\documentclass[a4paper,onecolumn,3p,preprint]{elsarticle}

\usepackage{lineno,hyperref}
\modulolinenumbers[5]
\usepackage[usenames,dvipsnames,svgnames,table]{xcolor}

\journal{Physica E}

\begin{document}

\title{Electronic transport in disordered graphene superlattices with scale-free correlated barrier spacements}

\author[UFRPE]{Anderson L. R. Barbosa}
\ead{anderson.barbosa@ufrpe.br}

\author[UFRPE,INT]{Jonas R. F. Lima}
\ead{jonas.lima@ufrpe.br}

\author[UFRPE]{\'Icaro S. F. Bezerra}
\ead{icarosfb.is@gmail.com}

\author[UFRPE,UFAL]{Marcelo L. Lyra}
\ead{marcelo@if.ufal.br}

\address[UFRPE]{Departamento de F\'{\i}sica, Universidade Federal Rural de Pernambuco, 52171-900, Recife, PE, Brazil}

\address[INT]{Institute of Nanotechnology, Karlsruhe Institute of Technology, D-76021 Karlsruhe, Germany}

\address[UFAL]{Instituto de F\'isica, Universidade Federal de Alagoas, 57072-970, Macei\'o - AL, Brazil}

\cortext[cor1]{Corresponding authors}

\date{\today}

\begin{abstract}
A transfer matrix approach is used to study the electronic transport in graphene superlattices with long-range correlated barrier spacements. By considering the low-energy electronic excitations as massless Dirac fermions, we compute by transmission spectra of graphene superlattices with potential barriers having spacements randomly distributed with long-range correlations governed by a power-law spectral density $S(k)\propto 1/k^{\alpha}$. We show that at large incidence angles, the correlations in the disorder distribution do not play a significant role in the electronic transmission. However, long-range correlations suppress the Anderson localization as normal incidence is approached and a band of transmitting modes sets up reminiscent of Klein tunneling.
\end{abstract}

\maketitle


\section{Introduction}

The very peculiar band structure of graphene with conductance and valence bands touching at special Fermi points with a linear dispersion relation around them allows to describe the low-energy electronic excitations as massless Dirac fermions\cite{dirac1,RevModPhys.81.109}. Such effective massless relativistic quasi-particle behavior is responsible for several unique electronic transport properties that have been explored in the engineering of new Graphene-based electronic devices\cite{dirac2,dirac3,dirac4,dirac5,dirac6,dirac7}.

There is a current wide research interest in graphene deposited in substrates due to the possibility of tailoring its electronic transport properties by the creation of barriers that can be controlled by external electric and magnetic fields\cite{field1,field2,field3,field4,field5,field6,field7,field8,field9,field10}. The transport of relativistic Dirac fermions through barriers displays the intriguing Klein tunneling paradox\cite{klein1,klein2,klein3,klein4}: Due to the spinorial character of the Dirac wavefunction it can be fully transmitted by an energy barrier higher than the particle's energy. This is in direct contrast with the scattering of non-relativistic quantum particle for which the transmission probability decays exponentially when the barrier height is enhanced.

Electronic transport in graphene superlattices have been a subject of increasing interest because they provide a powerful platform to manipulate both the spacial arrangement as well as the physical parameters of barriers\cite{super1,super2,super3,super4,super5}. In particular, the electronic transport through disordered one-dimensional graphene supperlattices can be strongly suppressed due to the Anderson localization phenomenon\cite{anderson1,anderson2,anderson3}. In one-dimensional channels with uncorrelated disorder, the one-particle eigenstates are exponentially localized which implies in the exponential damping of the transmitted wave as it traverses successive barriers. Dimer-like \cite{dimer1,dimer2,dimer3,dimer4,dimer5} and long-range\cite{long1,long2,long3,long4,long5,long6,long7,long8} correlations in the disorder distribution have been shown to be able to suppress Anderson localization in low-dimensional systems.

Deeply understanding of the interplay between Klein tunneling and Anderson localization in disordered graphene superlattices is fundamental to reach a complete description of their transport properties. It is well known that the localization length of massive relativistic Dirac particles in one-dimension is always larger than that of the corresponding non-relativistic particles. Further, massless relativistic particles remain entirely delocalized due to Klein tunneling\cite{ka1}. 
In a one-dimensional graphene supperlattice, the incidence angle of electrons plays a significant role in the transmission properties\cite{ka2,ka33}. The transmission
of quasiparticles with large incidence angles is suppressed by uncorrelated disorder in the width of the barriers due to Anderson localization. On the other hand, transmission still results reminiscent of Klein tunneling at small incidence angles. Recently, it was shown that these two regimes leads to either standard or anomalous scaling of the transmission when the barrier widths are distributed following a scale-free Levy distributions\cite{ka3}. Further, fractal\cite{frac}, self-affine potentials\cite{ka4} and long-range scale-free correlations in the barrier heights and widths\cite{ka5} as well in the velocity profile\cite{ka6} were also considered. By using the
trace of a fractional Brownian motion with a power spectrum $S(k)\propto 1/k^{\alpha}$ to generate a random long-range correlated sequence, it was shown that the conductance increases with increasing correlation-exponent\cite{ka5,ka6}. A transition to a conducting phase takes place at a critical correlation exponent which depends strongly on the disorder strength and rather weakly on the energy of the incident particles. However, studies on how long-range scale-free correlations in the spacial distribution of identical barriers  influence the electronic transport properties of graphene supperlattices is still missing. 

In the present work, we advance in the study of electronic transport in long-range correlated disordered graphene superlattices. We will use a transfer matrix technique to compute the average transmission of plane waves with a general incident angle $\theta$ and energy $E$ on a graphene sheet with a random distribution of potential barriers. Assuming the effective Dirac equation for massless fermions and a scale-free distribution of barrier spacements with power-law spectral density, we will obtain the spectra of the average transmission and its associated average logarithmic  as a function of the typical spectral exponent. A finite-size scaling analysis will be employed to emphasize the distinct transmission regimes induced by the scale-free character of the disorder.   

\section{Graphene superlattices with correlated barrier spacements}

\begin{figure}
\centering
\includegraphics[width=0.9\linewidth]{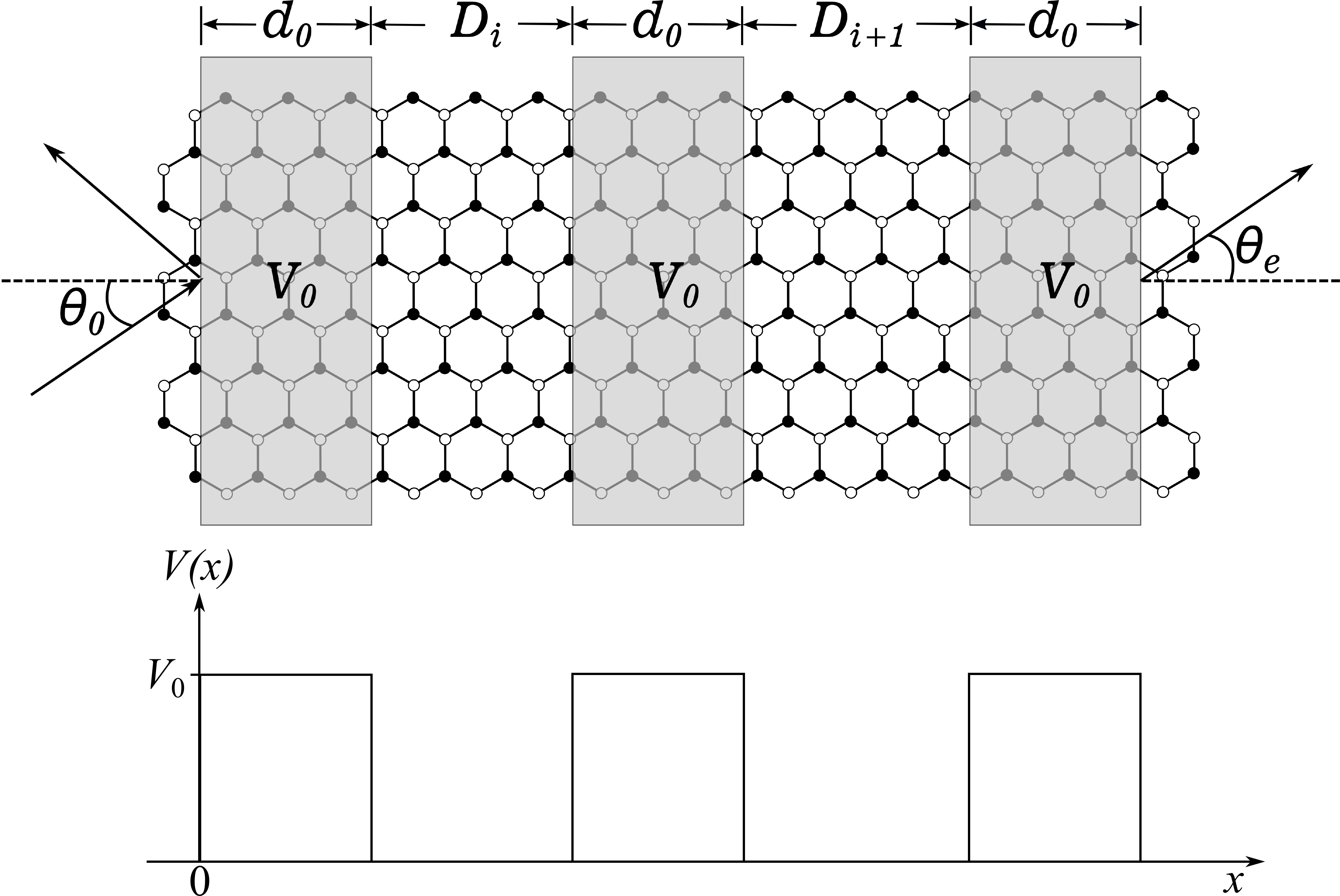}
\caption{Schematic diagram of the disordered graphene superlattice with correlated barrier spacements. }
\label{graphene}
\end{figure}

\begin{figure}
\centering
\includegraphics[width=1\linewidth]{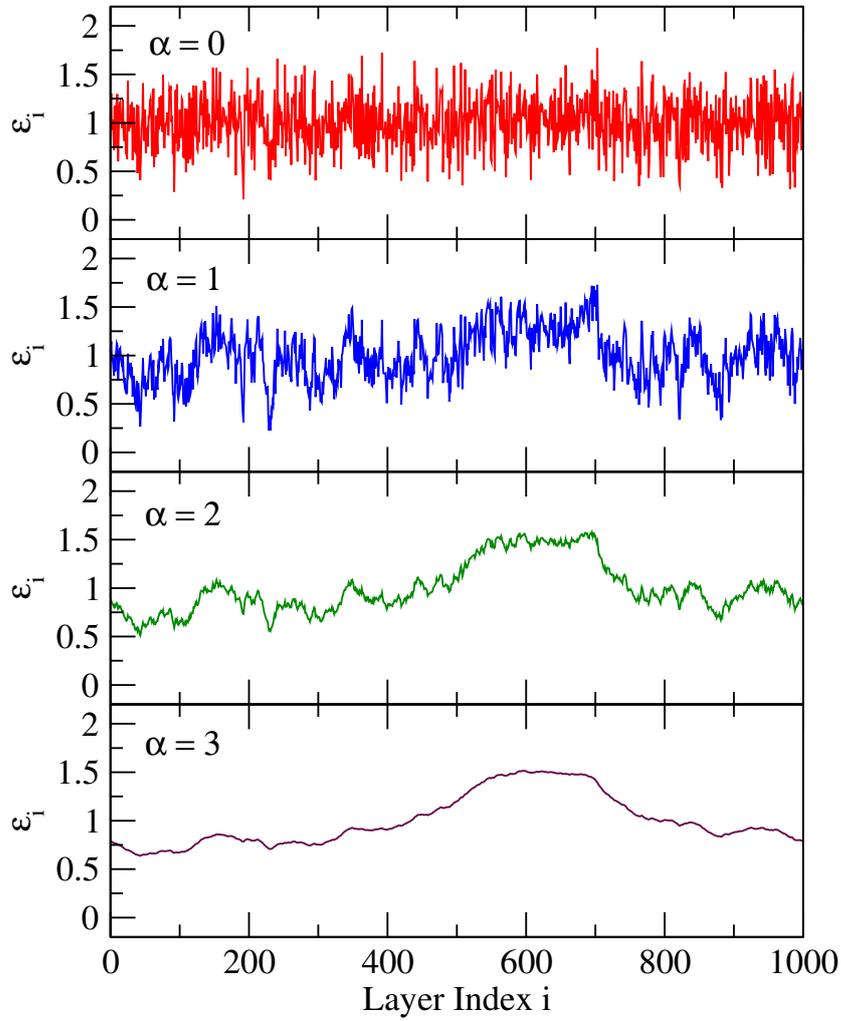}

\caption{The long-range correlated sequences generated by Eq. (\ref{lc}) after the proper displacement and normalization of the average and variance (see text). A few representative values of the spectral exponents  $\alpha=0,1,2$ and 3 (top to bottom) were used to generate sequences with $N=1000$.}
\label{Imagem0}
\end{figure}

We will consider the electronic transport in a graphene single layer with a one-dimensional potential pattern. Stripped barriers with electrostatic potential $V_0$ and width $d_0$ are randomly separated by regions with no electrostatic potential. A schematic representation is shown in Fig. \ref{graphene}.  The spacement between the potential barriers are randomly distributed encompassing long-range correlations.   We use a discrete Fourier transform to construct long-range correlated
sequences of $N$ barrier spacements having power-law spectral density by construction  given by\cite{long1,corr1,corr2,corr3,corr4}
\begin{equation}
\varepsilon_i=\sum_{k=1}^{N/2} \cos{\left(\frac{2\pi ik}{N}+\phi_k\right)}/k^{\alpha/2} ,   \label{lc} 
\end{equation}
where $\phi_i$ are random phases distributed uniformly in the interval $[0,2\pi]$.
The above sequence has power-law spectral density $S_k\propto 1/k^{\alpha}$ and corresponds to the trace of a fractional long-range correlated Brownian motion. In order to keep all values of this sequence positive, we displaced it uniformly to a minimum value $\varepsilon = 1$ with no impact on its spectral properties. After that, all values are scaled to have unitary average and the same standard deviation for all values of the spectral exponent $\alpha$. An illustration of the  long-range correlated sequences generated by the above scheme is given in Fig. (\ref{Imagem0}) for some representative values of the spectral exponent. The sequence of barrier spacements is chosen to be given by $D_i=d_0\varepsilon_i$. 

It is important to mention that we will consider here a single electron through the graphene superlattice, neglecting, this way, the electron-electron interaction. However, the many-body problem introduces only a renormalization on the Fermi velocity for the weak-coupling regime, which means that, depending on the electron's concentration, we should only change the value of $v_F$ \cite{Neto2012}. So, the single-electron problem reproduces qualitatively the many-body problem, which allows us not to take into account the Coulomb interaction. Furthermore, several experimental realizations of graphene superlattices with different potential barriers patterns were already reported in the literature \cite{PhysRevB.89.115421,PhysRevB.97.195410,doi:10.1063/1.4807888,10.1038/ncomms3342,KrishnaKumar181,PhysRevLett.121.036802}. Thus, the present model system can, in principle, be realizable following recently developed  experimental techniques.


Previous works analyzing disordered graphene superlattices, as Ref.  \cite{super1,super2,super3,super4,super5}, developed all calculations without considering the presence of long-range correlations in the barriers and/or barriers spacement, which means that the distribution of barriers was always similar to the one shown in Fig. (\ref{Imagem0}) for $\alpha = 0$. Our aim in this work is to go beyond $\alpha = 0$, which means to study the effect of long-range correlations in the barriers spacement, as shown in Fig. (\ref{Imagem0}) for $\alpha = 1,2$ and 3 (strong long-range correlations). Hence, we will show that, in the presence of sufficiently long-ranged correlations, the Anderson localization is suppressed and one sustains a transmission band reminiscent of Klein tunneling.

\section{Transport of massless Dirac fermions}

Since we have a piecewise constant electrostatic potential $V(x)$, which alternates between the values $V(x)=V_0$ and $V(x)=0$, we can use the transfer matrix method to obtain the transmittance of the system directly. The effective Dirac equation for the system is given by
\begin{eqnarray}
-i\hbar v_F\left(\sigma_x \partial_x +\sigma_y \partial_y\right)\psi  = [E-V(x)] \psi,
\label{diraceq}
\end{eqnarray}
where $\sigma_i$ are the Pauli matrices, $v_F$ is the Fermi velocity and $\psi=(\psi_A, \psi_B)^T$. Due to translational invariance in the $y$ direction, we can write $\psi(x,y)=e^{-ik_yy}\psi(x)$. So, inside the $j$th region in which $V(x)$ is constant, we obtain
\begin{equation}
\frac{d^2\psi_{A,B}}{dx^2} + (k^2_j - k^2_y) \psi_{A,B} = 0  ,
\end{equation}
where $k_j = (E-V_j)/(\hbar v_F)$ is the wave vector inside that region. The subscript $j$ denotes the regions of the system, $j = 0,1,2,...,2N+1,e$, where $j=0$ is the incident region, $j=e$ the exit region and $N$, as in the previous section, is the number of spacements between the barriers, which means that the number of barriers is $N+1$. The solutions are of the form
\begin{equation}
    \psi_{A,B}=C_{A,B}e^{iq_jx}+D_{A,B}e^{-iq_jx},
\end{equation}
where $C_{A,B}$ and $D_{A,B}$ are constants and $q_j$ is the $x$ component of the wave vector  given by $q_j = \sqrt{k_j^2-k_y^2}$.

The transfer matrix connecting the wave function $\psi(x)$ at $x$ and $x+\Delta x$ in the $j$th region is given by \cite{Yao10}
\begin{equation}
M_j(\Delta x, E, k_y)=\left(
\begin{array}{cc}
\frac{\cos(q_j\Delta x - \theta_j)}{\cos\theta_j} & i\frac{\sin(q_j\Delta x)}{\cos\theta_j} \\
i\frac{\sin(q_j\Delta x)}{\cos\theta_j} & \frac{\cos(q_j\Delta x + \theta_j)}{\cos\theta_j}
\end{array} \right) \; ,
\end{equation}
where $\theta_j$ is given by $\theta_j = \arcsin (k_y/k_j)$. Hence, the transfer matrix connecting incident and exit wave functions is given by $X = \prod_{j=1}^{2N-1} M_j(w_j, E, k_y)$, where $w_j$ is the width of the regions, which is equal to $d_0$ and $D_i$ for a region with $V(x)=0$ and $V(x)=V_0$, respectively. Note that the subscript $j$ labels the regions in the superlattice while the subscript $i$ labels the electrostatic barriers. The transmission coefficient is given by
\begin{equation}
t(E, k_y) = \frac{2\cos \theta_0}{(x_{22}e^{-i\theta_0}+x_{11}e^{-i\theta_e})-x_{12}e^{i(\theta_e-\theta_0)}-x_{21}} ,
\label{tc}
\end{equation}
where $x_{mn}$ are the matrix elements of $X$ and $\theta_0 (\theta_e)$ is the incidence (exiting) angle.

\section{Results and Discussion} 

\begin{figure}
\centering
\includegraphics[width=0.9\linewidth]{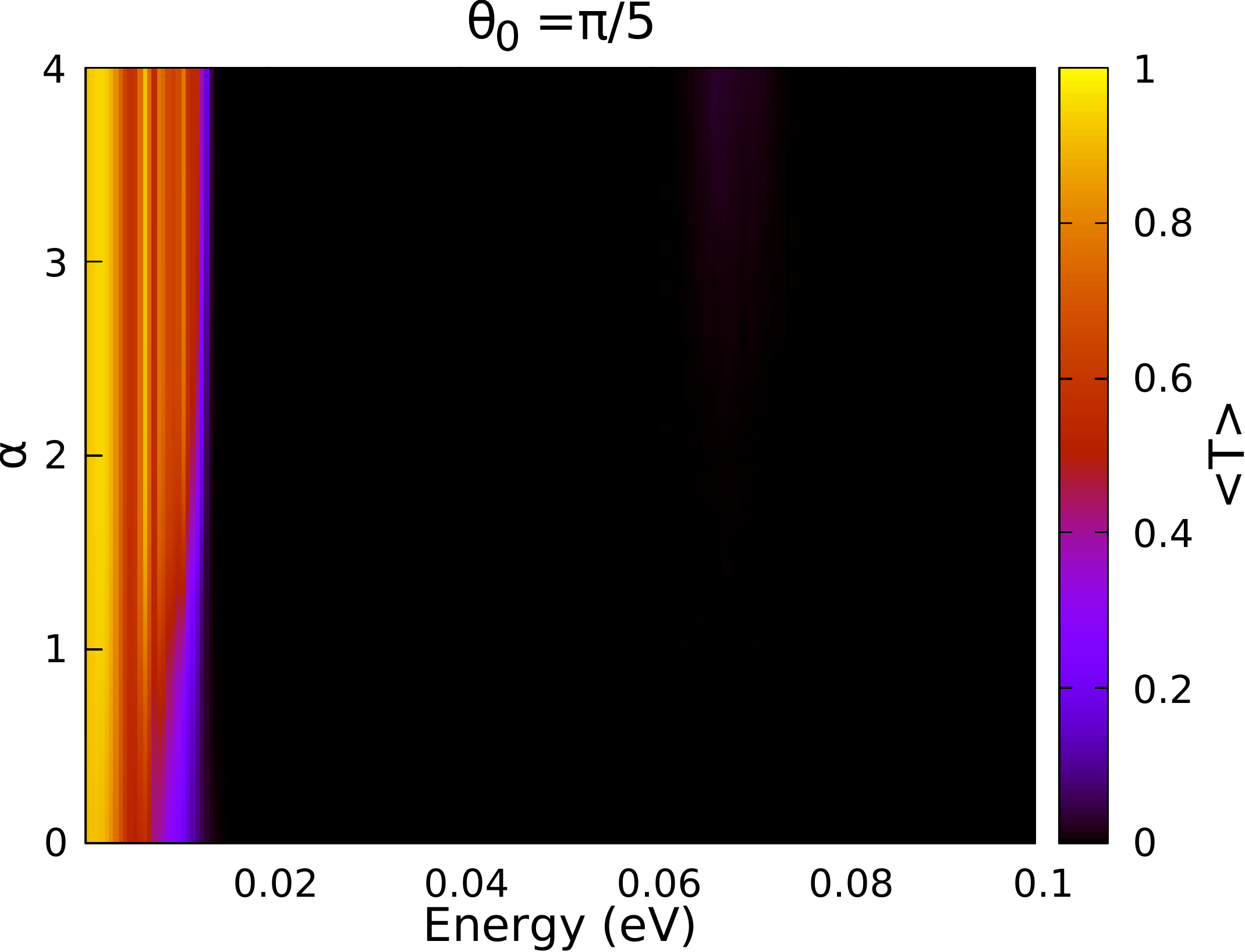}
\includegraphics[width=0.9\linewidth]{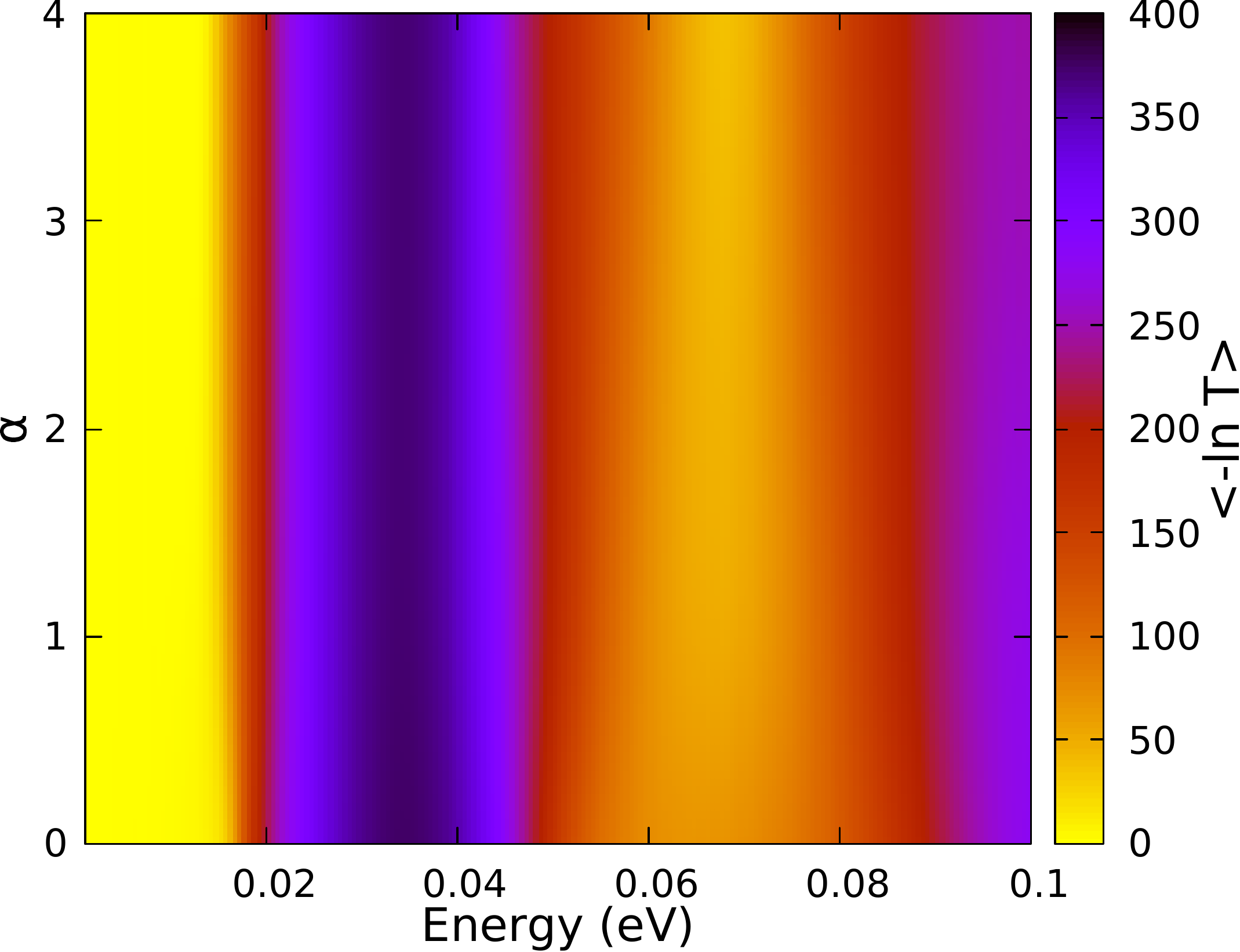}
\caption{Average transmission probability $\left\langle T \right\rangle$ (top panel) and the average logaritmic transmission $\langle-\ln{T}\rangle$ (bottom panel) as a function of waveparticle energy $E$ and the correlation exponent $\alpha$ for an incidence angle $\theta_0=\pi/5$ and $N=200$. Notice that the transmission spectra is roughly independent of the correlation exponent.}
\label{Imagem1}
\end{figure}

\begin{figure}
\centering
\includegraphics[width=0.9\linewidth]{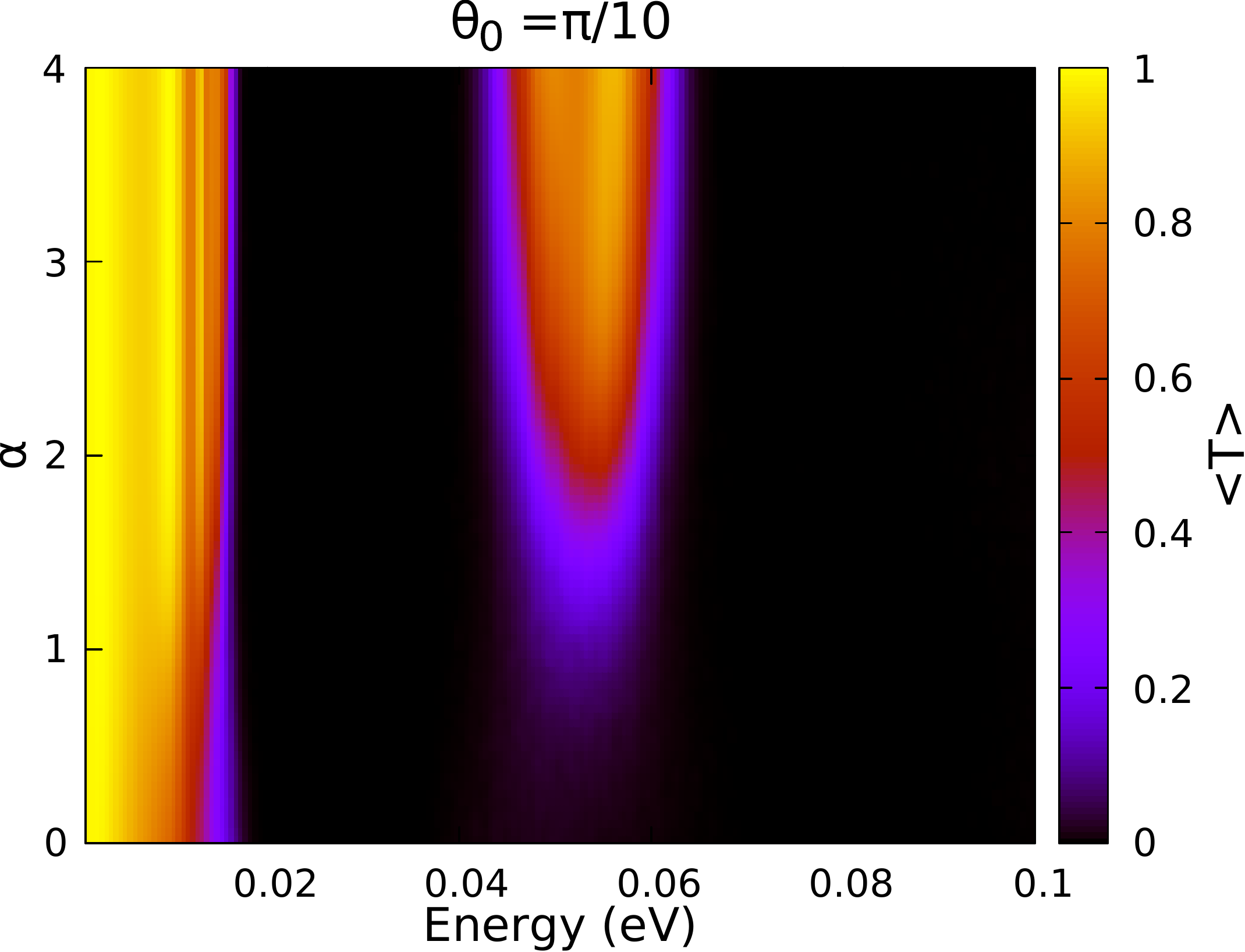}
\includegraphics[width=0.9\linewidth]{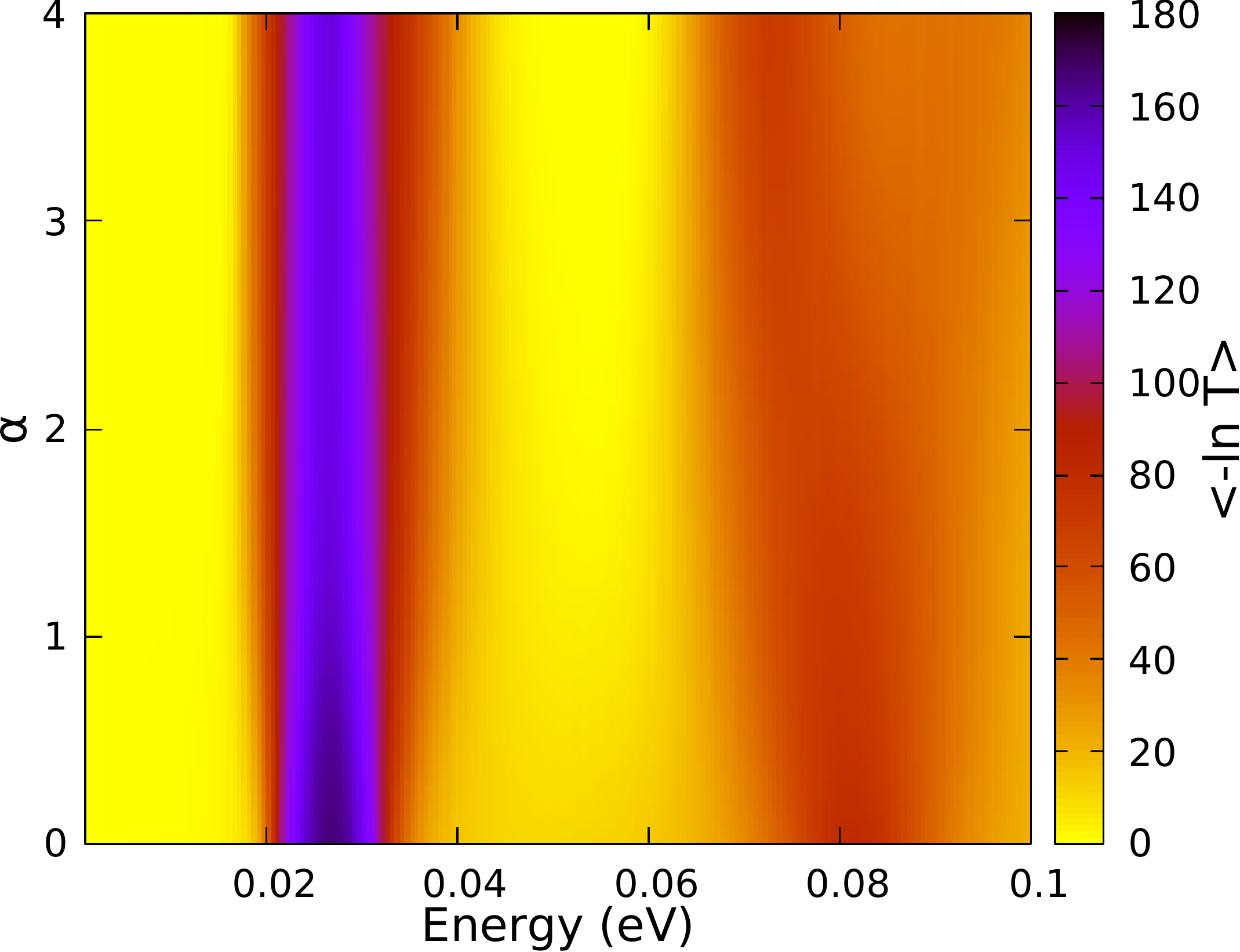}
\caption{Average transmission probability $\left\langle T \right\rangle$ (top panel) and the average logaritmic transmission $\langle-\ln{T}\rangle$ (bottom panel) as a function of waveparticle energy$E$ and the correlation exponent $\alpha$ for an incidence angle $\theta_0=\pi/10$ and $N=200$. Notice the emergence of a transmission band for large correlation exponents around the resonance condition $E=V_0=50$ meV.}
\label{Imagem2}
\end{figure}

In this section we provide numerical data regarding the electronic transport in a binary disordered graphene superlattice with long-range correlations is the distribution of the widths of the alternating regions. We start by reporting the transmission probability and its logarithmic derivative averaged over distinct disorder configurations as a function of the quasiparticle energy $E$ and the correlation exponent $\alpha$. In all numeric calculation was used an electrostatic potential $V_0 = 50$ meV, a width $d_0 = 20$ nm and Fermi velocity $v_F = 10^6$ m/s, while the transmission average was taken from $10^3$ realizations. 

It is very important to stress here that we are considering a one-dimensional graphene superlattice. As such, we are analyzing the transmission only in the $x$ direction, although graphene is a two-dimensional material. Then, the word ``localisation" here does not mean full localisation of the electronic states, but only localisation in the direction of the graphene superlattice, since the wave function remains delocalised in the $y$ direction. It is very well known that full Anderson localisation in graphene can only be achieved breaking the valley degeneracy \cite{Gonz_lez_Santander_2013,Bardarson_2007,Nomura_2007}, which is not done here.

\begin{figure}
\centering
\includegraphics[width=0.9\linewidth]{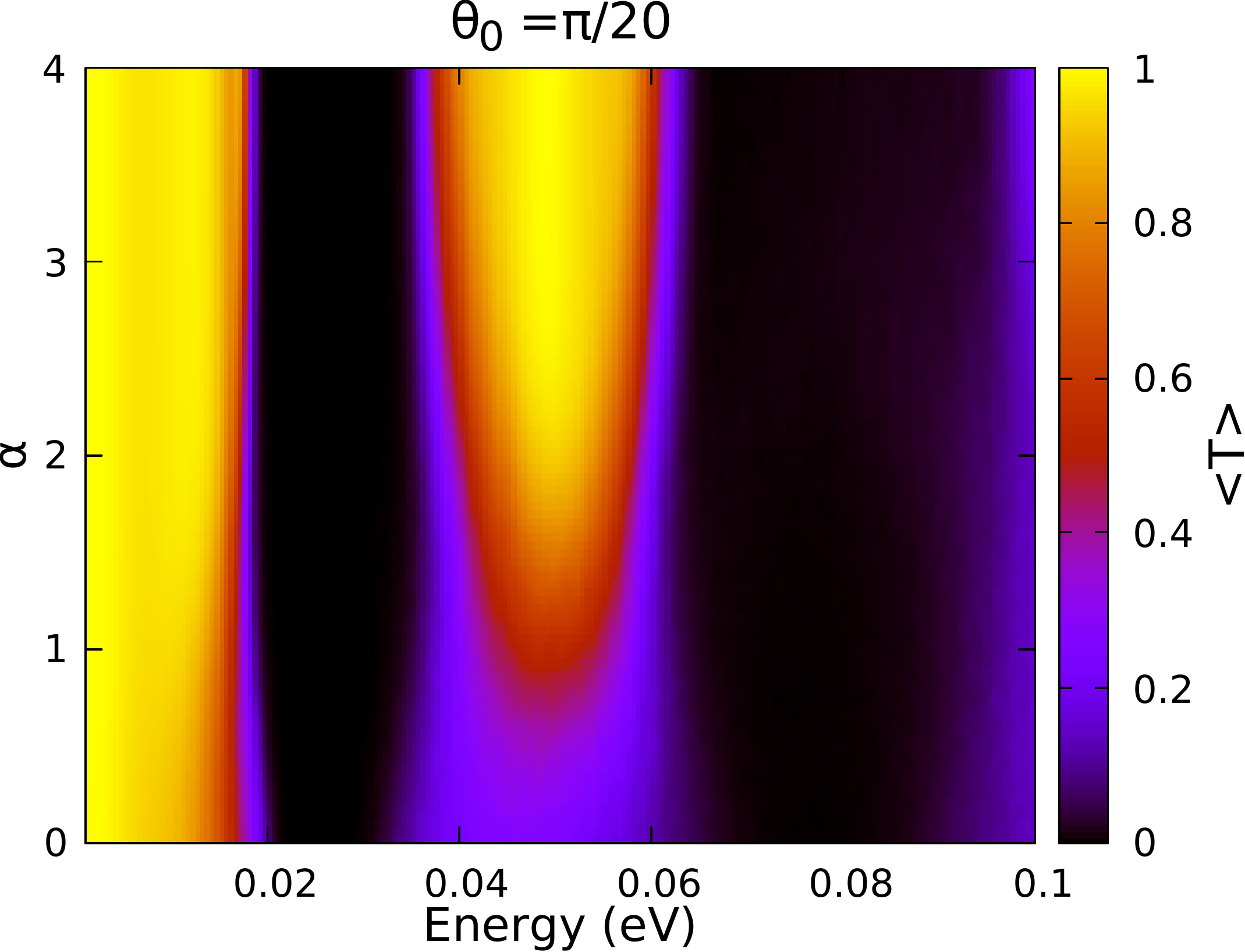}
\includegraphics[width=0.9\linewidth]{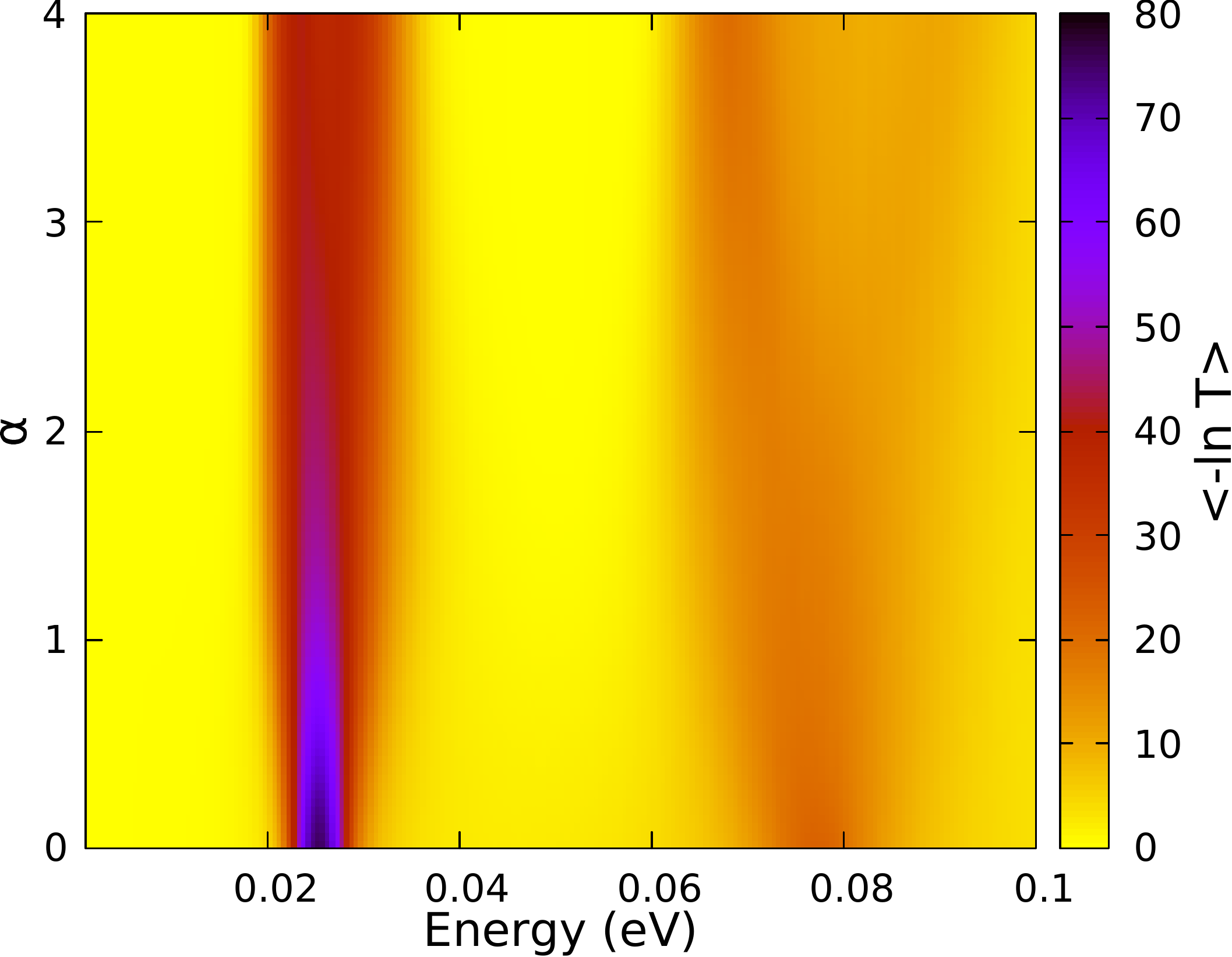}
\caption{Average transmission probability $\left\langle T \right\rangle$ (top panel) and the average logarithmic transmission $\langle-\ln{T}\rangle$ (bottom panel) as a function of waveparticle energy $E$ and the correlation exponent $\alpha$ for an incidence angle $\theta_0=\pi/20$ and $N=200$. The correlation-induced transmission band becomes more pronounced at low incidence angles for which an almost perfect Klein tunneling occurs.
}
\label{Imagem3}
\end{figure}

\begin{figure}
\centering
\includegraphics[width=1\linewidth]{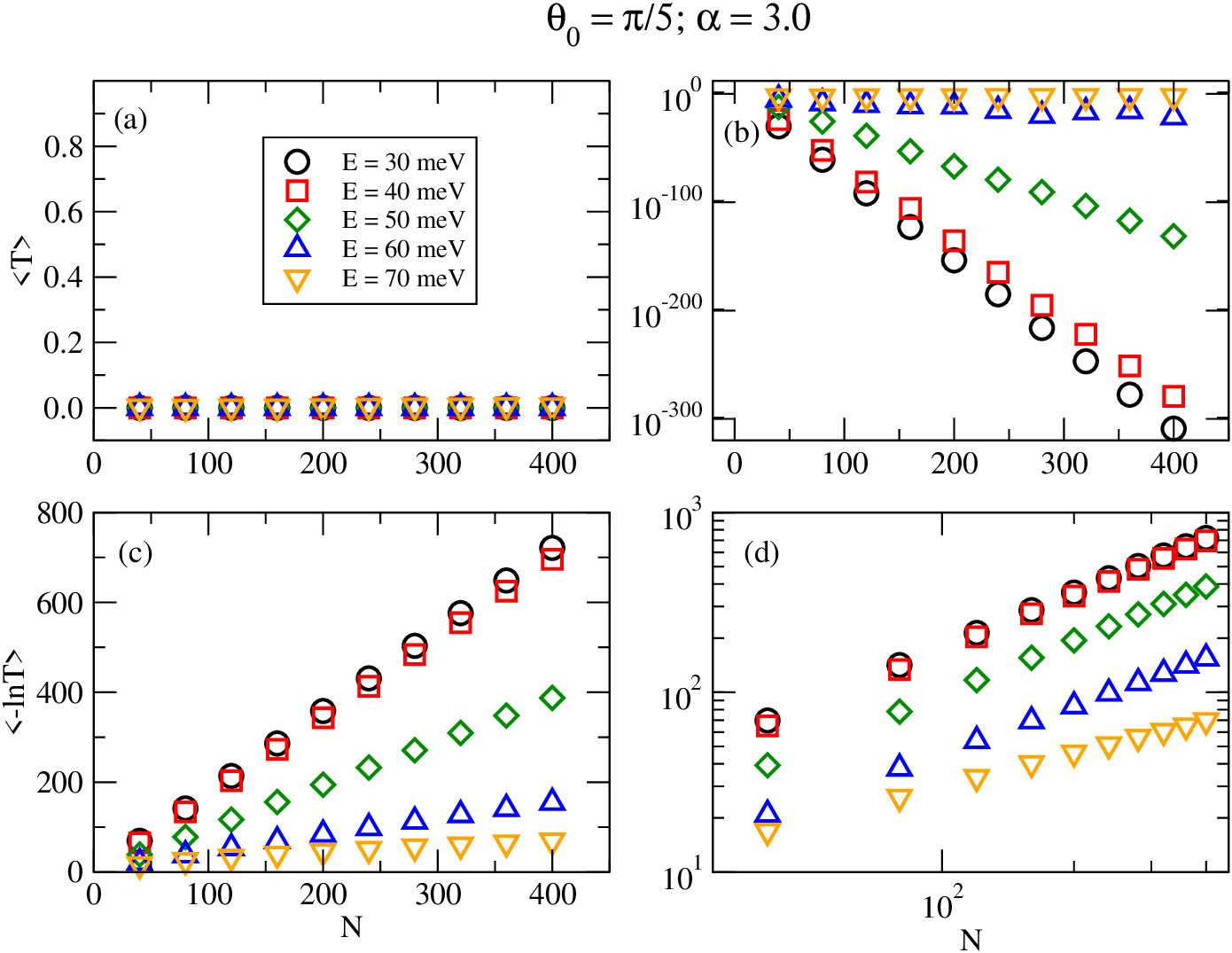}
\caption{Finite-size dependence of the average transmission (a-b) and average logarithmic transmission (c-d) for $\alpha=3$, $\theta_0=\pi/5$ and distinct quasiparticle energies. Linear (left) and logarithmic (right) scales are shown to emphasize the distinct scaling behaviors. Notice that the exponential localization becomes weaker at higher energies.}
\label{Imagem4}
\end{figure}

In Fig. (\ref{Imagem1}) we consider the case of a large incidence angle $\theta_0 = \pi/5$ and $N=200$. At large incidence angles, the transmission spectrum is fairly independent of the degree of long-range correlation present in the disorder distribution. Transmission is low, except at small energies. This feature indicates that correlations in the disorder distributions is not a relevant issue in this regime. 
The remaining transmission at low energies is reminiscent of the finite-size of the superlattice structure. In this regime, the  transmission through a single barrier is large and decays slowly at it goes through a sequence of barriers. Notice that the average logarithmic transmission is also smaller at energies of the order of the barrier height (although slightly displaced to higher energies), signalling a weaker wavefunction  damping.

The above picture changes significantly when smaller incidence angles are considered. In Fig. (\ref{Imagem2}) we report the corresponding results for $\theta_0 = \pi/10$ and $N=200$. Notice that a band of transmitting modes appears for large correlation exponents around the resonant energy $E=V_0$. In this spectral region, the quasiparticle wavefunction is partially transmitted through a single barrier with an evanescent character\cite{ka2}. Transmission is suppressed in this regime by Anderson localization. However, strong long-range correlations are known to induce a localization-delocalization transition\cite{long1,long2,long3,long4,long5,long6,long7,long8}, thus leading to the emergence of the reported transmission band. Notice that the average logarithmic transmission is not an appropriate quantity  to clearly signal this delocalization transition, in agreement with the similar delocalization phenomenon reported in graphene superlattices with Levy-distributed barrier spacements\cite{ka3}.

For even smaller incidence angles, one approaches to the condition of quasi-perfect Klein tunneling. We illustrate this case in Fig. (\ref{Imagem3}) reporting the transmission spectrum for $\theta_0 = \pi/20$ and $N=200$. Notice now that Anderson localization is weaker due to such almost perfect tunneling. Some degree of transmission takes place for energies of the order of the barrier height even for small correlation exponents, resulting from a pronounced finite-size effect. Here again the averaged logarithmic transmission does not clearly capture the correlation-induced delocalization transition. However, it signals the stronger Anderson localization for weakly correlated disorder away from the resonance condition. In this work, we say that $\theta_0$ is a large incidence angle when it departs sufficiently from normal incidence so that no transmission occurs irrespective to the degree of potential correlations.  When the transmission energy band vanishes for all values of $\alpha$, as it can be seen in the Fig. (\ref{Imagem1}) for $\theta_0 = \pi/5$, it can be considered a large incidence angle. This feature holds for even larger incidence angles. However, for incidence angles smaller than $\theta_0 = \pi/5$, a transmission energy band emerges in the regime of strong correlations, as it can be seen in the Figs. (\ref{Imagem2}) and (\ref{Imagem3}) for $\theta_0 = \pi/10$ and $\pi/20$.

To unveil the finite-size scaling behavior of the transmission in the vicinity of the spectral region around the resonance condition $E=V_0$, we computed the average transmission and its associated average logarithmic transmission as a function of the number of barriers present in the graphene superlattice for distinct energy values. We focus in the case of correlation exponent $\alpha=3$ for which a delocalization transition takes place at low incidence angles, as indicated in the above analysis. The same values of the incidence angles will be explored to allow a closer comparison with the reported transmission spectra.
\begin{figure}
\centering
\includegraphics[width=1\linewidth]{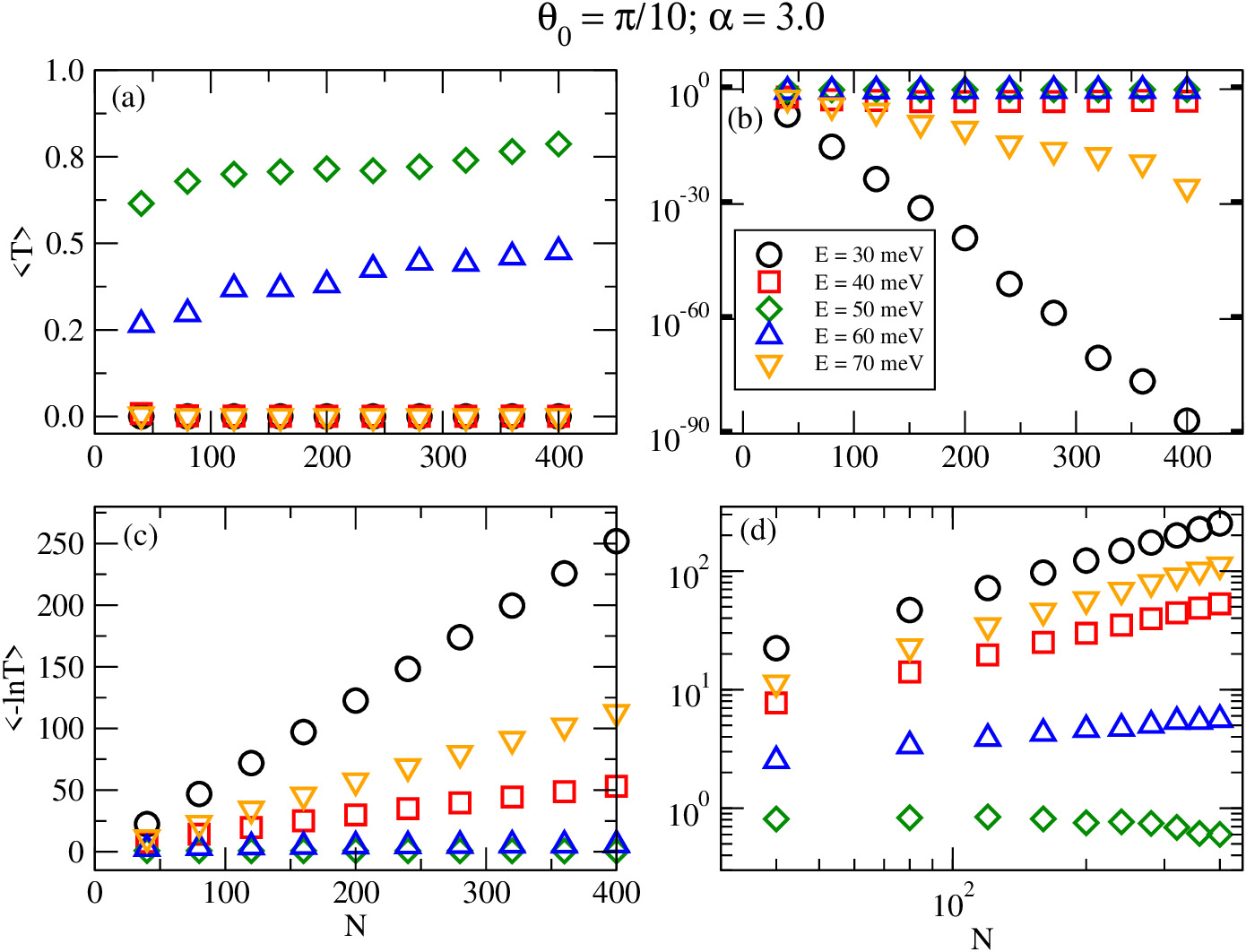}
\caption{Finite-size dependence of the average transmission (a-b) and average logarithmic transmission (c-d) for $\alpha=3$, $\theta_0=\pi/10$ and distinct quasiparticle energies. Linear (left) and logarithmic (right) scales are shown to emphasize the distinct scaling behaviors. Notice that these quantities become roughly size-independent at $E=V_0=50$ meV, supporting a correlation-induced suppression of Anderson localization. }
\label{Imagem5}
\end{figure}

In Fig. (\ref{Imagem4}) we consider the large incidence angle $\theta_0 =\pi/5$. Notice that the average transmission is quite small for all energies, as shown in Fig. (\ref{Imagem4}-a). When plotted in logarithmic scale (see Fig. \ref{Imagem4}-b) it exhibits the expected exponential decay which becomes slower as the quasiparticle energy grows. The average logarithmic transmission depicts a linear size dependence with the number of barriers, as reported in Fig. (\ref{Imagem4}-c), typical of exponential localization. This is better illustrated in the double log scale (see Fig. \ref{Imagem4}-d) with data depicting similar slopes, with a slow crossover for higher energies.

\begin{figure}
\centering
\includegraphics[width=1\linewidth]{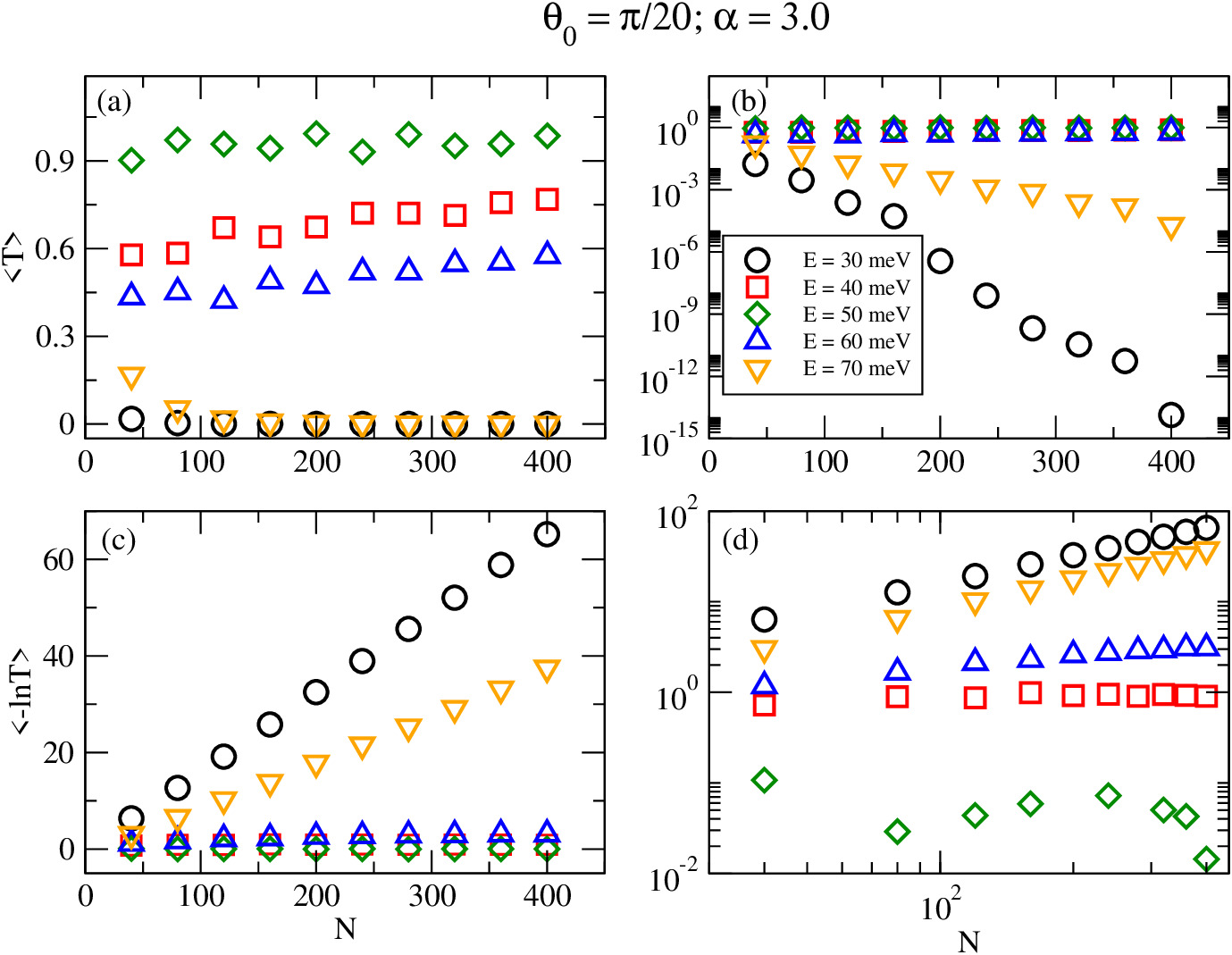}
\caption{Finite-size dependence of the average transmission (a-b) and average logarithmic transmission (c-d) for $\alpha=3$, $\theta_0=\pi/20$ and distinct quasiparticle energies. Linear (left) and logarithmic (right) scales are shown to emphasize the distinct scaling behaviors.
Correlation-induced delocalization takes place in the energy range $[40,60]$ meV. 
}
\label{Imagem6}
\end{figure}

The finite-size scaling of the average transmission for $\theta_0=\pi/10$ is summarized in Fig. (\ref{Imagem5}). Notice that the average transmission becomes roughly size independent for $E$ around $V_0=50$ meV, Fig. (\ref{Imagem5}-a). When plotted in logarithmic scale, one sees clearly the exponential decay with the system size for energies outside the transmission band. A similar trend is observed also in the finite-size scaling behavior of the average logarithmic transmission, Figs. (\ref{Imagem5}-c,d). In this case, we notice that for $E=60$ meV $\langle-\ln{T}\rangle$ grows quite slowly with the system size signalling that this energy is very close to the mobility edge. This feature can also be observed in Fig. (\ref{Imagem2}).  
Finally, the corresponding finite-size scaling for a smaller incidence angle $\theta_0=\pi/20$ is shown in Fig. (\ref{Imagem6}). The data supports the previous indication that the transmission band induced by the long-range correlations in the disorder distribution becomes wider. Both the average transmission and the average logarithmic transmission become mainly size-independent for energies in the range $[40,60]$ meV, in agreement with the transmission spectra shown in Fig. (\ref{Imagem3}). Note that, the value of $\alpha = 3$ was taken in Figs. (\ref{Imagem4}), (\ref{Imagem5}) and (\ref{Imagem6}) because it depicts a transmission energy band at low incidence angles, as shown in Figs. (\ref{Imagem1}), (\ref{Imagem2}) and (\ref{Imagem3}). The reported trend remains the same for larger values of $\alpha$.

The above features can be physically understood under the light of the competition between the Klein tunneling process that favors electron transmission near normal incidence and Anderson localization. In order to have a clear picture of the underlying physics, it is important to recall that perfect transmission through a single potential barrier occurs for the normal incidence. This Klein tunneling is not an interference effect between the two interfaces but rather due to the conservation of the pseudo-spin leading to the absence of backscattering\cite{ka2}. For oblique incidence some degree of backscattering is produced, which is weaker in the energy range around $V_0$ where the electron wavefunction has an evanescent character. For small incidence angles such backscattering is weak enough to allow for a finite transmission amplitude over the superlattices sizes investigated under the action of very long-range correlated disorder. For weakly correlated barrier sequences, Anderson localization predominates due to incoherent backscattering and the transmission quickly fades away. At energies far from the resonance condition $E=V_0$, strong incoherent backscattering leads to Anderson localization irrespective to the degree of correlations in the potential barriers distribution.

\begin{figure}
\centering
\includegraphics[width=0.8\linewidth]{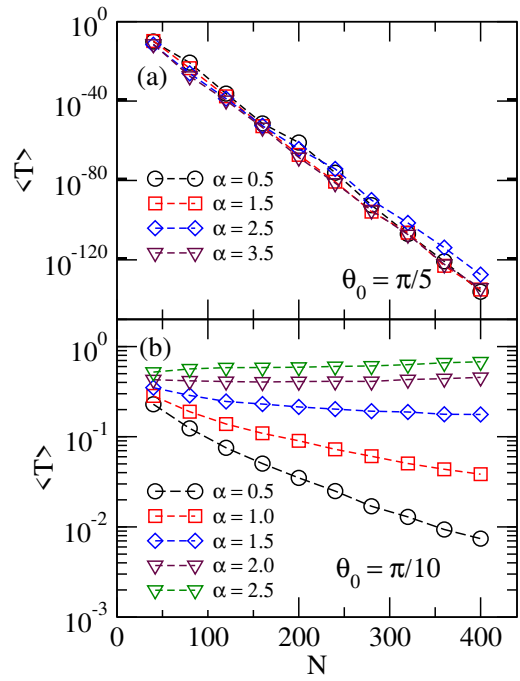}
\caption{Finite-size dependence of the average transmission for two representative incidence angles (a) $\theta_0 = \pi/5$ and (b) $theta_0 =\pi/10$ for a set of spectral exponents $\alpha$. Data are for $E=V_0=50$~meV. At large incidence angles, the average transmittance decays exponentially with the number of barrier spacements $N$, irrespective to the spectral exponent. On the other hand, it remains finite at low incidence angles for large values of $\alpha$.
}
\label{Imagem7}
\end{figure}

\begin{figure}
\centering
\includegraphics[width=0.9\linewidth]{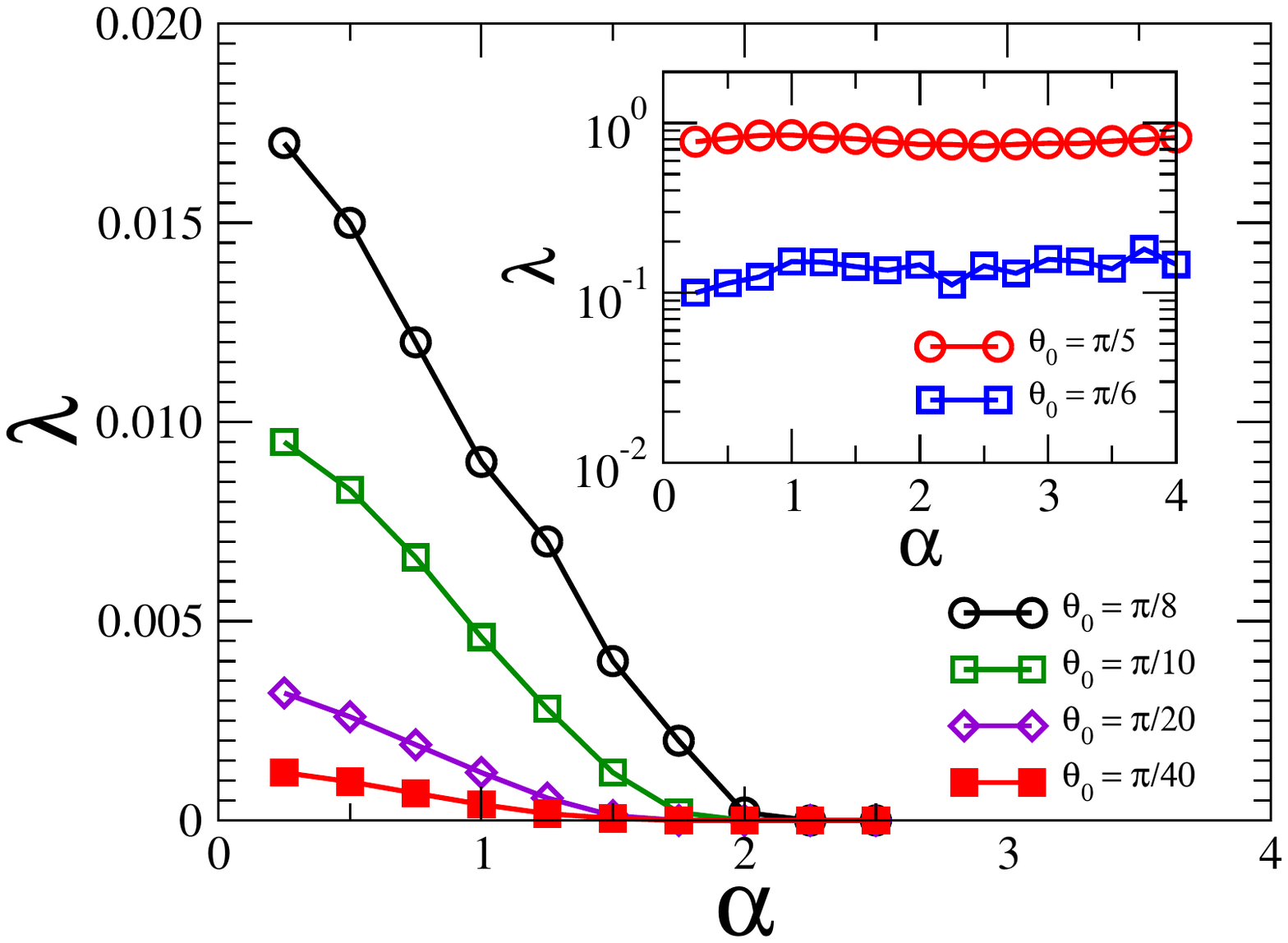}
\caption{Estimated Lyapunov coefficient (inverse localization length) as a function of the spectral exponent $\alpha$ for a set of incidence angles. Data are for $E=V_0=50$~meV. There is a clear localization-delocalization transition at $\alpha\simeq 2$  for small incidence angles. Above a characteristic incidence angle, the localization length remains finite for all values  of $\alpha$ (see inset). 
}
\label{Imagem8}
\end{figure}

A more detailed scenario regarding the influence of the spectral exponent $\alpha$ in the regimes of small and large incidence angles can be raised by investigating the finite-size scaling behavior of the average transmission at the resonance energy $E=V_0$. In Fig.(\ref{Imagem7}-a) we illustrate the case of a large incidence angle using $\theta_0=\pi/5$. Notice that the average transmittance decays exponentially with the number of barriers $N$, irrespective to the spectral exponent $\alpha$. However, the picture changes qualitatively at small incidence angles, as shown in Fig.(\ref{Imagem7}-b) for $\theta_0 = \pi/10$. In this case, the transmission remains finite as the number of barriers in enhanced whenever large values of the spectral exponent are considered.  Here, we can estimate the Lyapunov coefficient $\lambda$ by assuming that the asymptotic behavior of the transmission can be put in the form $\langle T\rangle\propto e^{-\lambda N}$. $\lambda$ is a measure of the inverse localization length. It remains finite when the transmission decays exponentially and vanishes when it converges to a constant value as $N\rightarrow \infty$.  In Fig.(\ref{Imagem8}) we summarize our results for the estimated values of $\lambda$ in a wide range of spectral exponents and incidence angles. Data from the average transmission at $E=V_0$ were fitted to the above exponential decay-law in the range $N=[100,400]$. A clear transition from a non-transmitting (finite $\lambda$)  to a transmitting (vanishing $\lambda$) is developed below a characteristic incidence angle. This transition takes place at $\alpha\simeq 2$ and is already seem at incidence angles $\theta_0 \leq \pi/8$. For larger incidence angles the Lyapunov coefficient remains finite irrespective to the spectral exponent $\alpha$. On the other hand,  the Lyapunov coefficient becomes vanishingly small at very small incidences due to the proximity of the perfect Klein tunelling condition, even for weakly correlated barrier distributions (small $\alpha$ values).

Fig. (\ref{Imagem7}-a) shows that for a large incidence angle ($\theta_0 = \pi/5$) the average transmittance decays exponentially (Anderson localization) because the transmission energy band vanishes for all values of $\alpha$. For all larger incidence angles $\theta_0 \geq \pi/5$ we obtain a similar behavior. Furthermore, the results of Fig. (\ref{Imagem7}-b) show that  the average transmittance remains finite for the low incidence angle $\theta_0 = \pi/10$ and large values of $\alpha$ because of the emergence of a transmission band around the resonance energy $E=V_0$. This picture remains qualitatively the same for smaller incidence angles $\theta_0$.

\section{Summary and Conclusions}

In summary, we studied the electronic transport properties of an one-dimensional graphene superlattice composed of striped potential barriers with spacements being randomly chosen according to a scale-free distribution having spectral density decaying as a power-law with a characteristic exponent $\alpha$ that governs the long-range character of the underlying disorder correlations. Our findings go beyond the well known results concerning  the electronic transmission on graphene superlattices with uncorrelated disorder  in the barriers height and/or spacements \cite{super1,super2,super3,super4,super5} and deepens the current knowledge about the role played by long-range correlated disorder.

The low-energy electronic excitations were modelled as massless Dirac fermions. As such, they exhibit perfect Klein tunneling at normal incidence even in the presence of strong disorder and high potential barriers. For large incidence angles, the transmission is vanishing small irrespective to the presence of correlations in the disorder distribution, except at very low energies where finite-size effects are pronounced because the transmission through a single barrier is large. At small incidence angles, we evidenced that long-range correlations can suppress Anderson localization and sustain a transmission band reminiscent of Klein tunneling in an energy band centered at the potential barrier value. Therefore, the present results show that the interplay of Klein tunneling and Anderson localization in graphene superlattices with correlated disorder can be explored for the proposal of new graphene-based devices with engineered electronic transport properties. 

\section{acknowledgements}

This work was supported by CAPES (Coordena\c{c}\~ao de Aperfei\c{c}oamento de Pessoal
de N\'{\i}vel Superior), CNPq (Conselho Nacional de Desenvolvimento Cient\'{\i}fico e Tecnol\'ogico), FAPEAL
(Funda\c{c}\~ao de Apoio \`a Pesquisa do Estado de Alagoas), AvH (Alexander von Humboldt Foundation) and FACEPE (Funda\c{c}\~ao de Amparo \`a Ci\^encia e Tecnologia do Estado de Pernambuco). M.L.L. acknowledges the hospitality of the Physics Department of Federal Rural University of Pernambuco where this work has been developed with partial financial support from a partnership program CAPES/FACEPE (grant number APQ-0325-1.05/18).

\end{document}